\documentclass{aa}  
\usepackage{natbib}
\bibpunct{(}{)}{;}{a}{}{,} 

\usepackage{graphicx}
\usepackage{txfonts}
\newcommand{\omegat}{\tilde{\omega}}
\newcommand{\wa}{\omega_{\rm A}}

\newcommand{\fb}{f_{\rm B}}
\newcommand{\gb}{g_{\rm B}}

\newcommand{\vai}{v_{\mathrm{Ai}}}
\newcommand{\ra}{r_{\mathrm{A}}}

\newcommand{\der}{{\rm d}}
\newcommand{\pd}{\partial}

\newcommand{\da}{\Delta_{\rm A}}
\newcommand{\drea}{\Lambda_{{\rm R}}}
\newcommand{\dima}{\Lambda_{{\rm I}}}

\newcommand{\vk}{v_{\rm k}}

\newcommand{\rhoi}{{\rho_{\rm i}}}
\newcommand{\rhon}{{\rho_{\rm n}}}

\newcommand{\ld}{L_{\rm D}}

\begin{document}

	\title{Resonant Alfv\'en waves in partially ionized\\ plasmas of the solar atmosphere}

	\titlerunning{Resonant waves in partially ionized plasmas}

   \author{R. Soler, J. Andries,  \and M. Goossens}
\offprints{R. Soler}
\institute{Centre for Plasma Astrophysics, Department of Mathematics, Katholieke Universiteit Leuven,
              Celestijnenlaan 200B, 3001 Leuven, Belgium \\
                       \email{roberto.soler@wis.kuleuven.be} }

 	 \date{Received XXX / Accepted XXX}

  \abstract
   {Magnetohydrodynamic (MHD) waves are ubiquitous in the solar atmosphere. In magnetic waveguides resonant absorption due to plasma inhomogeneity naturally transfers wave energy from large-scale motions to small-scale motions. In the cooler parts of the solar atmosphere as, e.g., the chromosphere, effects due to partial ionization may be relevant for wave dynamics and heating.}
   {We study resonant Alfv\'en waves in partially ionized  plasmas. }
   {We use the multifluid equations in the cold plasma approximation.  We investigate propagating resonant MHD waves in partially ionized flux tubes. We use approximate analytical theory based on normal modes in the thin tube and thin boundary approximations along with numerical eigenvalue computations.}
   {We find that the jumps of the wave perturbations across the resonant layer are the same as in fully ionized plasmas. The damping length due to resonant absorption is inversely proportional to the frequency, while that due to ion-neutral collisions is inversely proportional to the square of the frequency.  For observed frequencies in the solar atmosphere, the amplitude of MHD kink waves is more efficiently damped by resonant absorption than by ion-neutral collisions. }
   {Most of the energy carried by chromospheric kink waves is converted into localized azimuthal Alfv\'en waves that can deposit energy in the coronal medium. The dissipation of wave energy in the chromosphere due to ion-neutral collisions is only effective for high-frequency waves. The chromosphere acts as a filter for kink waves with periods shorter than 10~s.}


   \maketitle


\section{Introduction}

Recent observations have shown the ubiquitous presence of propagating magnetohydrodynamic (MHD) Alfv\'enic waves in the solar atmosphere. For example, Alfv\'enic transverse waves propagating in magnetic waveguides of the solar corona were first observed by \citet{tomczyk07} and \citet{tomczyk09} using the Coronal Multichannel Polarimeter (CoMP), and more recently by \citet{mcintosh2011} using SDO/AIA. In chromospheric spicules, the presence of Doppler oscillations is known for more than 40 years \citep[see the review by][]{temuryreview}. Recent observations of Alfv\'enic transverse waves in spicules have been reported by, e.g, \citet{depontieu07}, \citet{temuryspicules}, \citet{kim}, \citet{he1,he2}, \citet{okamotodepontieu}. In addition, propagating transverse waves in thin threads of solar prominences have been observed \citep[e.g.,][]{lin07,lin09} and Alfv\'enic waves in bright points have been reported \citep{jess2009}. The role and implications of the observed  Alfv\'enic waves for the heating of the solar atmosphere have been discussed by, e.g., \citet{robertus07,depontieu07,mcintosh2011,cargillineke}.

Based on MHD wave theory, a number of works have interpreted the observed waves as propagating kink MHD waves \citep[e.g.,][]{robertus07,tom08,lin09,pascoe1,pascoe2,TGV,VTG,verthspicule,solerspatial,resonantflow,stratified}. Kink waves are transverse waves with mixed fast MHD and Alfv\'enic properties \citep[see, e.g.,][]{edwinroberts,goossens09}. In thin magnetic tubes kink waves are highly Alfv\'enic because their dominant restoring force is magnetic tension \citep{goossens09}. It has been shown that resonant absorption, caused by plasma inhomogeneity in the direction transverse to the magnetic field, is a natural and efficient damping mechanism for kink waves \citep[see the recent reviews by][]{goossens06,goossens08,goossens11}. In magnetic waveguides resonant absorption transfers wave energy from transverse kink motions to azimuthal motions localized in the inhomogeneous part of the waveguide. This process has been studied numerically by \citet{pascoe1,pascoe2} in the case of driven kink waves in coronal  waveguides.

Using analytical theory based on the thin tube and thin boundary approximations, \citet[hereafter TGV]{TGV} obtained that the damping length due to resonant absorption is inversely proportional to the wave frequency. Therefore, it was predicted that high-frequency kink waves become damped in length scales shorter than low-frequency waves. \citet{VTG} showed that this result is consistent with the CoMP observations of damped coronal waves \citep{tomczyk07,tomczyk09}. Subsequent investigations have extended the original work by  TGV by incorporating effects not considered in their paper. For example, \citet{resonantflow} took the presence of flow into account, and \citet{stratified} studied the influence of longitudinal density stratification. Both works concluded that the damping length  remains inversely proportional to the frequency when flows and longitudinal stratification are present.

In TGV and subsequent works cited above, the plasma is assumed fully ionized. However, the plasma in the cooler parts of the solar atmosphere is only partially ionized  as, e.g., in the chromosphere or in prominences. This fact raises the relevant question on whether the damping length remains inversely proportional to the frequency when the plasma is partially ionized or, on the contrary, this dependence is modified by the effect of ion-neutral collisions. The effect of partial ionization on the damping of Alfv\'en waves has been investigated in a large number of papers in different contexts. Some  examples are the works by, e.g., \citet{hearendel,depontieu98,pecseli,depontieu2001,forteza07,solerpartial,marc2010,temury,temuryhelium}. However, these works  only focus on the role of partial ionization for the damping and do not consider  resonant absorption, which is a basic and unavoidable phenomenon when plasma and/or magnetic  inhomogeneity is present. 

To our knowledge, the first attempt to study resonant waves in partially ionized plasmas was by \citet{solerpartialres}. These authors used the single-fluid approximation \citep[see, e.g.,][]{brag} to investigate standing resonant kink waves in a model of a partially ionized prominence thread. Subsequently, this first investigation was extended in \citet[hereafter SOB]{solerspatial} to the case of propagating waves. The use of the single-fluid approximation, as in  \citet{solerpartialres} and SOB, seems reasonable as the expected values of the collision frequencies in the solar atmosphere are much larger than the observed wave frequencies. However, the single-fluid approximation misses important effects when small length scales and/or high frequencies are involved. In such cases, the multifluid description is a more suitable approach \citep[see, e.g.,][]{temury,temuryhelium}. For resonant waves, perturbations develop very small length scales in the vicinity of the resonance position \citep[see, e.g.,][]{tirry,rudermanwright,vasquez,terradas2006}, where multifluid effects may play a relevant role.

Here we perform a general description of propagating resonant Alfv\'enic waves in partially ionized plasmas using the multifluid treatment. Our work extends the investigation of SOB by considering arbitrary values of the collision frequencies and by performing a more in-depth analysis of the resonant process. Section~\ref{sec:basic} contains a description of the equilibrium configuration and the basic equations. In Section~\ref{sec:resonant} the behavior of wave perturbations around the Alfv\'en resonance position are investigated and jump relations for the perturbations are derived. Later, in Section~\ref{sec:tube} we perform an application to resonant kink waves in straight tubes and obtain expressions for the damping lengths due to resonant absorption and ion-neutral collisions. The approximate analytical theory is complemented with fully numerical eigenvalue computations. In Section~\ref{sec:chromos} the implications of our results for the particular case of chromospheric waves is discussed. Finally, we give our main conclusions in Section~\ref{sec:con}.

\section{Equilibrium and basic equations}
\label{sec:basic}

We consider a partially ionized  plasma composed of ions, electrons, and neutrals. We use cylindrical coordinates, namely $r$, $\varphi$, and $z$ for the radial, azimuthal, and longitudinal coordinates. The medium is permeated by an equilibrium magnetic field, $\bf B$, of the form
\begin{equation}
 {\bf B} = B_\varphi \hat{e}_\varphi + B_z \hat{e}_z.
\end{equation}
In general, both azimuthal, $B_\varphi$, and longitudinal, $B_z$, components are functions $r$. The equations governing the dynamics of a magnetized multifluid plasma are discussed in classical works as, e.g., \citet{cowling} and \citet{brag}, or more recently in \citet{birk} and \citet{pinto}. Recent investigations which make extensive use of the multifluid description in the context of MHD waves are \citet{temury,temuryhelium}. We refer the reader to these two papers for details about the derivation of the equations. In brief, a partially ionized multifluid plasma is governed by the equations of the different species, which contain terms that couple the various fluids by means of collisions.

Here we study linear perturbations superimposed on the equilibrium state. Due to the very small momentum of electrons, we neglect the collisions of electrons with ions and neutrals. In addition, we consider a magnetically dominated plasma and neglect the gas pressure of ions compared to the Lorentz force, and the gas pressure of neutrals compared to the collisional friction with ions. This simplification neglects the plasma displacement along the magnetic field direction and removes the slow or cusp continuum from the equilibrium. Thus, in the present work we focus on resonance absorption in the Alfv\'en continuum only. The inclusion of gas pressure is a subject for future works. Under these conditions, the basic equations of our investigation are  
\begin{eqnarray}
 \rhoi \frac{\pd {\bf v}_{\rm i}}{\pd t} &=& \frac{1}{\mu} \left( \nabla \times {\bf b} \right) \times {\bf B} - \rhoi \rhon \gamma_{\rm in} \left( {\bf v}_{\rm i} - {\bf v}_{\rm n} \right), \label{eq:momion} \\
\rhon \frac{\pd {\bf v}_{\rm n}}{\pd t} &=&  - \rhoi \rhon \gamma_{\rm in} \left( {\bf v}_{\rm n} - {\bf v}_{\rm i} \right), \label{eq:momneu} \\
 \frac{\pd {\bf b}}{\pd t} &=& \nabla \times \left({\bf v}_{\rm i} \times {\bf B}  \right), \label{eq:induct}
\end{eqnarray}
 where ${\bf v}_{\rm i}$ and ${\bf v}_{\rm n}$ are the velocities of ions and neutrals, respectively, ${\bf b}$ is the magnetic field perturbation, $\rhoi$ and $\rhon$ are the densities of the ion and neutral fluids, respectively, $\mu$ is the magnetic permittivity, and $\gamma_{\rm in}$ is the ion-neutral collision rate coefficient per unit mass. Instead of using $\gamma_{\rm in}$, in the remaining of this paper we use the ion-neutral collision frequency, $\nu_{\rm in}$, which has a more obvious physical meaning. The ion-neutral collision frequency is defined as
\begin{equation}
 \nu_{\rm in} = \rho_{\rm i} \gamma_{\rm in}.
\end{equation}
Equations~(\ref{eq:momion}) and (\ref{eq:momneu}) are the linearized momentum equations of ions and neutrals, respectively. Equation~(\ref{eq:induct}) is the linearized induction equation, in which we have omitted all nonideal, diffusion terms as, e.g., magnetic resistivity. We do so because the goal of the present paper is to assess the particular role of ion-neutral collisions on the resonance absorption process. The effects of other dissipative mechanisms as, e.g, resistivity or viscosity, have been extensively studied in the existing literature \citep[see the recent review by][]{goossens11}

In this work the equilibrium quantities are functions of $r$ alone, so that the equilibrium is  uniform in both azimuthal and longitudinal directions. Hence we can write the perturbed quantities proportional to $\exp \left( i m \varphi + i k_z z  \right)$, where $m$ and $k_z$ are the azimuthal and longitudinal wavenumbers, respectively.  We express the temporal dependence of the perturbations as $\exp \left( - i \omega t \right)$, with $\omega$ the frequency. Since there are no equilibrium flows, the Lagrangian displacement of ions, ${\bf \xi}_{\rm i}$, and neutrals, ${\bf \xi}_{\rm n}$, are 
\begin{equation}
 {\bf \xi}_{\rm i} = \frac{i}{\omega} {\bf v}_{\rm i}, \qquad  {\bf \xi}_{\rm n} = \frac{i}{\omega} {\bf v}_{\rm n}.
\end{equation}

From Equation~(\ref{eq:momneu}) it is straightforward to derive the relation between ${\bf \xi}_{\rm n}$ and ${\bf \xi}_{\rm i}$, namely
\begin{equation}
 {\bf \xi}_{\rm n} = \frac{i\nu_{\rm in} }{\omega + i\nu_{\rm in} } {\bf \xi}_{\rm i}. \label{eq:xiixin}
\end{equation}
Equation~(\ref{eq:xiixin}) informs us that ${\bf \xi}_{\rm n}$ and ${\bf \xi}_{\rm i}$ are related to each other. The factor of proportionally is a complex function of the wave frequency, $\omega$, and the ion-neutral collision frequency, $\nu_{\rm in}$. Hence, there is a phase difference between the motions of ions and neutrals. Physically, this phase difference can be interpreted as the ``time delay'' between the motions of ions and the corresponding reaction of  neutrals. Thus, for $\nu_{\rm in} \gg \omega$, ${\bf \xi}_{\rm n} = {\bf \xi}_{\rm i}$ and the phase difference is zero, i.e., ions and neutrals behave as a single fluid. On the contrary, for $\nu_{\rm in} \ll \omega$ ions and neutrals become decoupled from each other, i.e., the collisionless case. In the remaining of this paper, we call the limits $\nu_{\rm in} \gg \omega$ and $\nu_{\rm in} \ll \omega$  the single-fluid limit and the collisionless limit, respectively.

Although we perform a general analysis for arbitrary  $\nu_{\rm in}$ and $\omega$, we must note that $\nu_{\rm in} \gg \omega$ is the realistic situation according to the observed frequencies of MHD waves in the solar atmosphere and the expected values of the collision frequency \citep[see, e.g.,][]{depontieu98, depontieu2001}. Thus, the case $\nu_{\rm in} \gg \omega$ will receive special attention.

\section{General expressions for resonant Alfv\'en waves}
\label{sec:resonant}

In this section we study the properties of resonant MHD waves in the Alfv\'en continuum.  We follow the notation used in, e.g., \citet{SGH91,goossens92,goossens95}. We combine Equations~(\ref{eq:momion})--(\ref{eq:induct}) to arrive at two coupled equations for the radial component of the Lagrangian displacement of ions, $\xi_{{\rm i}r} = i {\bf v_{{\rm i}}} \cdot \hat{e}_r / \omega$, and the Eulerian perturbation of the total pressure, $P={\bf B} \cdot {\bf b} / \mu$, namely 
\begin{eqnarray}
 \mathcal{D} \frac{\der  \left( r \xi_{{\rm i}r}  \right)}{\der r} &=& \mathcal{C}_1 r \xi_{{\rm i}r} - \mathcal{C}_2 r P, \label{eq:bas1} \\
 \mathcal{D} \frac{\der P}{\der r}  &=& \mathcal{C}_3 \xi_{{\rm i}r} - \mathcal{C}_1 P,\label{eq:bas2}
 \end{eqnarray}
with
\begin{eqnarray}
 \mathcal{D} &=& \rhoi \vai^2 \left( \omegat^2 - \wa^2  \right), \label{eq:defd}\\
\mathcal{C}_1 &=& \frac{2}{\mu} \frac{B_\varphi^2}{r} \omegat^2 - \vai^2 \frac{2}{\mu} \frac{m}{r^2} \fb B_\varphi, \label{eq:c1} \\
\mathcal{C}_2 &=&  \omegat^2 - \vai^2 \left( \frac{m^2}{r^2} + k_z^2 \right), \label{eq:c2} \\
\mathcal{C}_3 &=& \mathcal{D} \left[ \rhoi \vai^2 \left( \omegat^2 - \wa^2  \right) + \frac{2}{\mu} B_\varphi \frac{\der}{\der r} \left( \frac{B_\varphi}{r} \right) \right] \nonumber \\
              &+& \frac{4}{\mu^2} \frac{B_\varphi^4}{r^2}\omegat^2 - \frac{4}{\mu} \rhoi \vai^2 \wa^2 \frac{B_\varphi^2}{r^2} \label{eq:c3},
\end{eqnarray}
where $\vai^2 = \frac{B^2}{\mu \rhoi}$ is the square of the Alfv\'en velocity of the ions, with $B^2 = B_\varphi^2  + B_z^2$. In addition we have defined
\begin{eqnarray}
 \fb &=& \frac{m}{r}B_\varphi + k_z B_z, \\
\wa^2 &=& \frac{1}{\mu \rhoi} \fb^2, \\
\omegat^2 &=& \omega^2 \frac{\omega + i \left( 1 + \alpha \right) \nu_{\rm in}}{\omega + i\nu_{\rm in}}, \label{eq:omegat} \\
\alpha &=& \frac{\rhon}{\rhoi}.
\end{eqnarray}
In these expressions, $\wa$ is the local Alfv\'en frequency for the ions, $\omegat$ is the effective or modified frequency due to collisions, and $\alpha$ indicates the plasma ionization degree.

\subsection{Modified Alfv\'en continuum}

Equations~(\ref{eq:bas1}) and (\ref{eq:bas2}) are singular when $\mathcal{D} = 0$. As the equilibrium quantities are functions of $r$, the position of the singularity is mobile and depends on the frequency, $\omega$. The whole set of frequencies satisfying $\mathcal{D} = 0$ at some $r$ form a continuum of frequencies called the Alfv\'en continuum \citep[see, e.g.,][]{appert}. In the ideal, fully ionized case the condition $\mathcal{D} = 0$ is satisfied where the frequency, $\omega$, matches the local Alfv\'en frequency, $\wa$. 

We investigate how the Alfv\'en continuum is affected by ion-neutral collisions. The equation $\mathcal{D} = 0$, which $\mathcal{D}$ defined in Equation~(\ref{eq:defd}), can be rewritten as a third-order polynomial in $\omega$, namely
\begin{equation}
 \omega^3 + i \left( 1 + \alpha \right) \nu_{\rm in} \omega^2 - \wa^2 \omega - i \nu_{\rm in} \wa^2 = 0. \label{eq:continuum}
\end{equation}
Equation~(\ref{eq:continuum}) describes the Alfv\'en continuum modified by ion-neutral collisions. For $\nu_{\rm in} = 0$ we recover the ideal Alfv\'en continuum in a fully ionized plasma, i.e., $\omega = \pm \wa$. Note that Equation~(\ref{eq:continuum}) and so the continuum frequencies are independent of $m$. 

To shed light on the effect of ion-neutral collisions, it is convenient to rewrite Equation~(\ref{eq:continuum}) in the following form
\begin{equation}
\omega^2 - \wa^2 \frac{1 + i \frac{\nu_{\rm in}}{\omega} }{1 + i \left( 1 + \alpha \right) \frac{\nu_{\rm in}}{\omega}} = 0.
\end{equation}
This enables us to easily evaluate the continuum frequencies at the limits $\nu_{\rm in} \gg \omega$ and $\nu_{\rm in} \ll \omega$, namely
\begin{equation}
\omega \approx \left\{
\begin{array}{lll}
\pm \wa \left( 1+\alpha \right)^{-1}, & \textrm{for} & \nu_{\rm in} \gg \omega, \\
\pm \wa, & \textrm{for} & \nu_{\rm in} \ll \omega.
\end{array}  \right.
\end{equation}
In the limit $\nu_{\rm in} \ll \omega$ we obtain again the continuum frequencies of the ideal, collisionless case. In the limit $\nu_{\rm in} \gg \omega$, the continuum frequencies are modified by the plasma ionization degree but the frequencies remain real, so that the position of the singularity of Equations~(\ref{eq:bas1}) and (\ref{eq:bas2}) is shifted with respect to the collisionless case. 

The limits $\nu_{\rm in} \gg \omega$ and $\nu_{\rm in} \ll \omega$ are extreme cases. Now, we explore the continuum frequencies for arbitrary $\nu_{\rm in}$ to have a complete picture of the role of ion-neutral collisions. For arbitrary $\nu_{\rm in}$ we find an approximate solution to Equation~(\ref{eq:continuum})  by assuming that the effect of collisions on the continuum is to produce a weak damping of the continuum modes. This means that the continuum frequencies are assumed complex and that their real parts are taken the same as in the collisionless case. Thus we write $\omega \approx \pm \wa + i \gamma$, with $\gamma \ll \wa$, and put this expression in Equation~(\ref{eq:continuum}). After neglecting terms with $\mathcal{O} \left( \gamma^2  \right)$ we obtain the modified Alfv\'en continuum frequencies, namely
\begin{equation}
 \omega \approx \pm \wa - i \frac{1}{2} \alpha \nu_{\rm in}. \label{eq:continuum2}
\end{equation}
Equation~(\ref{eq:continuum2}) shows that the continuum frequencies in a partially ionized plasma are complex. Based on this result we anticipate that waves driven with a real frequency $\omega$ will not produce true singularities in  Equations~(\ref{eq:bas1}) and (\ref{eq:bas2}) since the Alfv\'en continuum frequencies are in the complex plane. This means that the singularity is removed in a partially ionized plasma even in the absence of additional dissipative mechanisms as, e.g., resistivity or viscosity.

As the two solutions given in Equation~(\ref{eq:continuum2}) are complex, the third solution to Equation~(\ref{eq:continuum}) must be purely imaginary. So, we write $\omega \approx i \gamma$ and neglect terms with $\mathcal{O} \left( \gamma^2  \right)$ to obtain an expression for the third solution, namely
\begin{equation}
 \omega \approx - i \nu_{\rm in},
\end{equation}
which corresponds to a purely damped, nonpropagating mode whose damping rate is given by the collision frequency. The presence of this third solution is not relevant for our subsequent analysis.

\subsection{Behavior of perturbations around the resonance position}

Here our purpose is to assess the behavior of the perturbed quantities around the position of the ideal Alfv\'en resonance. We denote by $\ra$ the ideal Alfv\'en resonance position. The value of $\ra$ is obtained from the ideal resonance condition, i.e., $\omega^2 = \wa^2 (\ra)$, where $\omega$ is assumed to be real. Around $r=\ra$ the behavior of the perturbations is dominated by the resonance. At $r=\ra$ ion-neutral collisions become relevant to remove the singularity of the coefficient $\mathcal{D}$ in Equations~(\ref{eq:bas1}) and (\ref{eq:bas2}), while their effect on the rest of coefficients is minor. This means that around $r=\ra$ we can take the ideal expressions of coefficients $\mathcal{C}_1$, $\mathcal{C}_2$, and $\mathcal{C}_3$, and only keep the terms related to collisions in the expression of  coefficient $\mathcal{D}$.

We define the new variable $s=r-\ra$. The ideal coefficients $\mathcal{C}_1$, $\mathcal{C}_2$, and $\mathcal{C}_3$ at $s=0$ become the following constant terms,
\begin{eqnarray}
 \mathcal{C}_1 &\approx& - \frac{2}{\mu} \frac{B_\varphi B_z}{\ra} \frac{\fb \gb}{\mu \rhoi}, \label{eq:c1s0} \\
\mathcal{C}_2 &\approx& - \frac{\gb^2}{\mu \rhoi},\label{eq:c2s0} \\
\mathcal{C}_3 &\approx& - \frac{4}{\mu^2} \frac{B_\varphi^2 B_z^2}{\ra^2} \frac{\fb^2}{\mu \rhoi},\label{eq:c3s0}
\end{eqnarray}
with $\gb = \frac{m}{\ra} B_z - k_z B_\varphi$. All quantities in Equations~(\ref{eq:c1s0})--(\ref{eq:c2s0}) have to be evaluated at $s=0$. Then we can rewrite Equations~(\ref{eq:bas1}) and (\ref{eq:bas2}) as
\begin{eqnarray}
 \mathcal{D}\frac{\der \xi_{{\rm i}r}  }{\der s} &\approx& \frac{\gb}{\mu \rhoi} \mathcal{C}_{\rm A}, \label{eq:bas1s0}\\
  \mathcal{D}\frac{\der  P}{\der s} &\approx& \frac{2}{\mu}   \frac{B_\varphi B_z}{\ra}\frac{\fb }{\mu \rhoi}\mathcal{C}_{\rm A}, \label{eq:bas2s0}\\
\mathcal{D}\frac{\der  \mathcal{C}_{\rm A}}{\der s} &\approx& 0,
\end{eqnarray}
where $ \mathcal{C}_{\rm A}$ is defined as
\begin{equation}
  \mathcal{C}_{\rm A} = \gb P - \frac{2}{\mu}   \frac{B_\varphi B_z}{\ra} \fb  \xi_{{\rm i}r}. \label{eq:conservation}
\end{equation}
The quantity $ \mathcal{C}_{\rm A}$ is approximately conserved across the resonance position. This is the same conservation law as that in ideal and dissipative MHD \citep[e.g.,][]{SGH91,goossens95,tirry}. The  Lagrangian displacement perpendicular to the magnetic field lines, $\xi_{{\rm i}\perp} = \left( \xi_{{\rm i}\varphi} B_z - \xi_{{\rm i}z} B_\varphi \right) / B$, is related to $ \mathcal{C}_{\rm A}$ as
\begin{equation}
\mathcal{D}  \xi_{{\rm i}\perp} = i \frac{B}{\mu \rhoi} \mathcal{C}_{\rm A}. \label{eq:xiperp}
\end{equation}

Next we approximate the coefficient $\mathcal{D}$ by its first-order Taylor polynomial around $s=0$.  The linear expansion of $\mathcal{D}$ is approximately valid in the interval $-s_{\rm A} <s < s_{\rm A}$, with $s_{\rm A}$ satisfying the relation $s_{\rm A} \ll \left| (\wa^2)' / (\wa^2)'' \right|$, where the prime denotes radial derivative. The resulting expression is
\begin{equation}
 \mathcal{D} \approx \rhoi \vai^2 \da \left( \Lambda + s  \right), \label{eq:ds0}
\end{equation}
with 
\begin{equation}
 \da = \frac{\der}{\der s} \left( \omega^2 - \wa^2 \right),
\end{equation}
and $\Lambda =  \drea + i \dima $, where $\drea$ and $\dima$ are the real and imaginary parts of $\Lambda$, respectively, given by
\begin{equation}
 \drea = \frac{\omega^2}{\da} \frac{\alpha \nu_{\rm in}^2}{\omega^2 + \nu_{\rm in}^2}, \qquad \dima = \frac{\omega^2}{\da} \frac{\alpha \omega \nu_{\rm in}}{\omega^2 + \nu_{\rm in}^2}.
\end{equation}
 As before all quantities in the previous expressions have to be evaluated at $s=0$. We find a constant complex term, $\Lambda$, in the expansion of $\mathcal{D}$ (Equation~(\ref{eq:ds0})). The presence of this term removes the singularity of the solutions, so that $\mathcal{D} \ne 0$ for $s=0$.   We use Equation~(\ref{eq:ds0}) to rewrite Equations~(\ref{eq:bas1s0}) and (\ref{eq:bas2s0}) as
\begin{eqnarray}
 \frac{\der \xi_{{\rm i}r}  }{\der s} &\approx& \frac{\gb}{\mu \rhoi^2 \vai^2} \frac{\mathcal{C}_{\rm A}}{\da} \frac{1}{\Lambda + s },\label{eq:bas1s02} \\
\frac{\der  P}{\der s} &\approx& \frac{2}{\mu}   \frac{B_\varphi B_z}{\ra}\frac{\fb }{\mu \rhoi^2 \vai^2} \frac{\mathcal{C}_{\rm A}}{\da}  \frac{1}{\Lambda + s }.\label{eq:bas2s02}
\end{eqnarray}
For our subsequent analysis it is convenient  to use the scaled variable $\tau$ defined as
\begin{equation}
 \tau = \frac{s + \drea}{\dima}.
\end{equation}
Then we integrate Equations~(\ref{eq:bas1s02}) and (\ref{eq:bas2s02}) and obtain
\begin{eqnarray}
 \xi_{{\rm i}r} &\approx&  \frac{\gb}{\mu \rhoi^2 \vai^2} \frac{\mathcal{C}_{\rm A}}{\da} \mathcal{G}(\tau) + \mathcal{C}_\xi, \\
P &\approx& \frac{2}{\mu}   \frac{B_\varphi B_z}{\ra}\frac{\fb }{\mu \rhoi^2 \vai^2} \frac{\mathcal{C}_{\rm A}}{\da} \mathcal{G}(\tau) + \mathcal{C}_P,
\end{eqnarray}
where $\mathcal{C}_\xi$ and $\mathcal{C}_P$ are constants of integration. The $\mathcal{G}(\tau)$ function is defined as
\begin{equation}
\mathcal{G}(\tau) \equiv \ln \left( \tau + i\right). \label{eq:psi}
\end{equation}
We separate the real and imaginary parts of $\mathcal{G}(\tau)$, namely
\begin{equation}
\Re \left( \mathcal{G}(\tau) \right)= \ln \sqrt{\tau^2 + 1}, \label{eq:psire}
\end{equation}
\begin{equation}
\Im \left( \mathcal{G}(\tau) \right)= \left\{
\begin{array}{lll}
  \arctan \tau^{-1} + \pi, & \textrm{for} & \tau < 0, \\
 \arctan \tau^{-1}, & \textrm{for} & \tau > 0. 
\end{array} \right. \label{eq:psiim}
\end{equation}
Note that we need to add $\pi$ to the imaginary part of $\mathcal{G}(\tau)$ for $\tau < 0$ in order to make the solution continuous and analytic at $\tau=0$. This is not a problem since we can always incorporate a constant of integration.

From Equation~(\ref{eq:xiperp}) we  obtain that the perpendicular displacement behaves as
\begin{equation}
 \xi_{{\rm i}\perp} \approx \frac{\mathcal{C}_{\rm A}}{\rhoi B } \frac{1}{\da \dima} \mathcal{F} (\tau),
\end{equation}
with $\mathcal{F}(\tau)$ defined as
\begin{equation}
 \mathcal{F}(\tau) \equiv \frac{i}{\tau + i}. \label{eq:phi}
\end{equation}
We separate the real and imaginary parts of $\mathcal{F}(\tau)$, namely
\begin{equation}
\Re \left( \mathcal{F}(\tau) \right)= \frac{1}{\tau^2+1}, \qquad \Im \left( \mathcal{F}(\tau) \right)= \frac{\tau}{\tau^2+1}. 
\end{equation}

\begin{figure}[!t]
\centering
 \includegraphics[width=0.95\columnwidth]{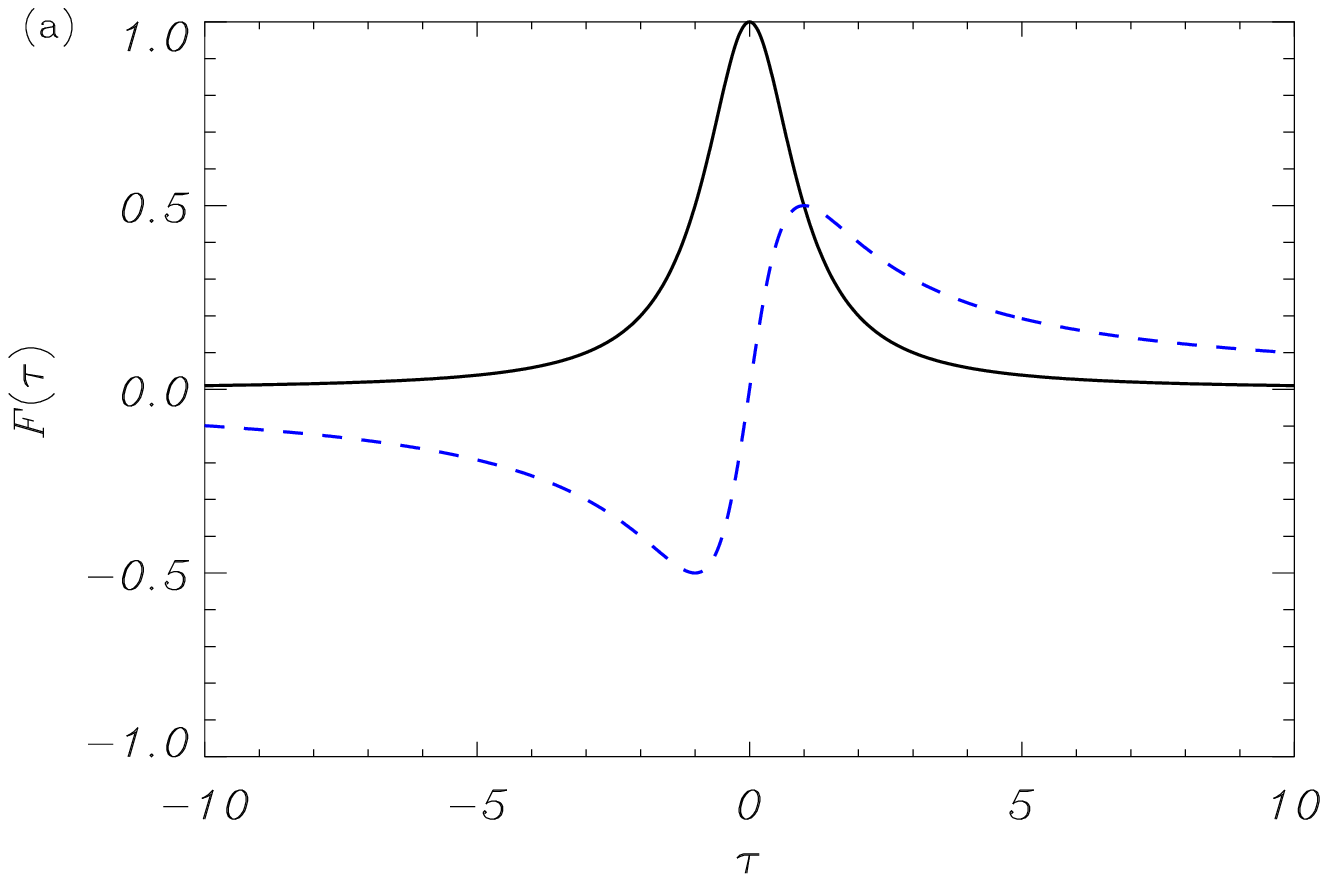}
 \includegraphics[width=0.95\columnwidth]{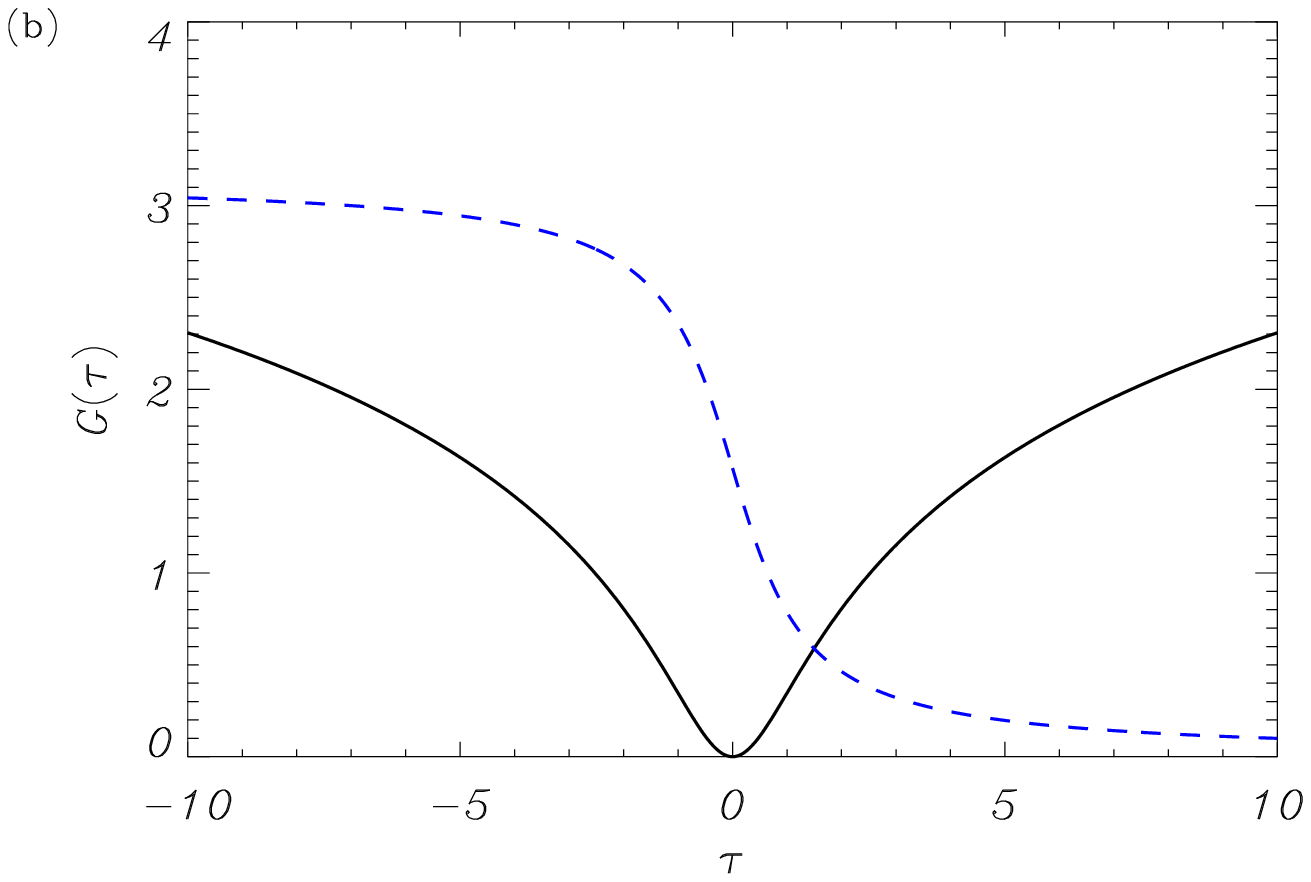}
\caption{Real (solid line) and imaginary (dashed line) of functions {\bf a)} $\mathcal{F}(\tau)$ and {\bf b)} $\mathcal{G}(\tau)$. \label{fig:functions}}
\end{figure}

Functions $\mathcal{F}(\tau)$ and $\mathcal{G}(\tau)$ play here the role of the {\em universal functions} found in a number of previous investigations of resonant waves in both stationary and non-stationary states and for different dissipative processes \citep[see, e.g.,][]{mok,goossens95,ruderman95,tirry,erdelyi,wrightallan,rudermanwright,vanlommel}. A comprehensive review on the importance and properties of the universal $\mathcal{F}(\tau)$ and $\mathcal{G}(\tau)$ functions can be found in \citet{goossens11}. In particular, our $\mathcal{F}(\tau)$ and $\mathcal{G}(\tau)$ functions coincide with  the functions found by \citet{wrightallan} for magnetospheric Alfv\'en waves damped by Pedersen conductivity (see their Equations~(23) and (24)), and are also equivalent to the functions described by \citet{vanlommel} for non-stationary quasi-modes in the cusp continuum (see their Equation~(16)). Figure~\ref{fig:functions} displays the real and imaginary parts of functions $\mathcal{F}(\tau)$ and $\mathcal{G}(\tau)$.

Ion-neutral collisions generate a {\em collisional layer} around the location of the resonance, i.e., $\tau = 0$, in a similar way as resistivity and/or viscosity generate a dissipative layer \citep[see, e.g.,][]{hollwegyang,poedts,SGH91}. The thickness of the collisional layer, $\delta$, is proportional to $\dima$, namely
\begin{equation}
 \delta \sim \left| \dima \right| = \frac{\omega^2}{\left| \da \right|} \frac{\alpha \omega \nu_{\rm in}}{\omega^2 + \nu_{\rm in}^2}. \label{eq:thicknu}
\end{equation}
It can be seen that $\delta \to 0$ in both limits $\nu_{\rm in} \gg \omega$ and $\nu_{\rm in} \ll \omega$, whereas $\delta$ takes its maximum value for $\nu_{\rm in} = \omega$. This result points out that the collisional layer is extremely thin for realistic collision frequencies, i.e,, $\nu_{\rm in} \gg \omega$.

\subsection{Jump Conditions}

Here we use the expressions derived in the last section to determine the jump of the perturbations across the collisional layer. In the scaled variable $\tau$ and assuming $\omega > 0$, the limits $s \to \pm s_{\rm A}$ are equivalent to the limits $\tau \to \pm \textrm{sign} (\da) \infty$. We denote by $[X]$ the jump of the quantity $X$ across the collisional layer, which we define as
\begin{equation}
 [X] = \lim_{\tau\to \textrm{sign} (\da)\infty} X (\tau) - \lim_{\tau\to -\textrm{sign} (\da)\infty} X(\tau). 
\end{equation} 
Hence, the jumps of $\xi_{{\rm i}r}$ and $P$ are
\begin{eqnarray}
 \left[\xi_{{\rm i}r}\right] &=&  \frac{\gb}{\mu \rhoi^2 \vai^2} \frac{\mathcal{C}_{\rm A}}{ \da } [\mathcal{G}], \\
\left[ P \right] &=&  \frac{2}{\mu}   \frac{B_\varphi B_z}{\ra}\frac{\fb }{\mu \rhoi^2 \vai^2} \frac{\mathcal{C}_{\rm A}}{\da }[\mathcal{G}].
\end{eqnarray}
To calculate the jumps of $\xi_{{\rm i}r}$ and $P$ we need the jump of $\mathcal{G}$, namely
\begin{equation}
 \left[\mathcal{G}\right] = - i \textrm{sign} (\da) \pi.
\end{equation}
Then it is straightforward to obtain
\begin{eqnarray}
 [\xi_{{\rm i}r}] &=& - i \pi \frac{\gb}{\mu \rhoi^2 \vai^2} \frac{\mathcal{C}_{\rm A}}{\left| \da \right|} , \label{eq:jumpxi} \\
\left[ P \right] &=& - i \pi  \frac{2}{\mu}   \frac{B_\varphi B_z}{\ra}\frac{\fb }{\mu \rhoi^2 \vai^2} \frac{\mathcal{C}_{\rm A}}{\left| \da \right|}. \label{eq:jumpp}
\end{eqnarray}
We find the remarkable result that these jumps are the same as obtained for ideal and dissipative MHD \citep[e.g.,][]{SGH91,goossens95,ruderman95}. This means that ion-neutral collisions do not modify the jumps of the plasma perturbations across the resonant layer. In the particular case of a straight magnetic field, i.e., $B_\varphi = 0$, the conserved quantity across the resonant layer is proportional to the total pressure perturbation, and the jump conditions for $\xi_{{\rm i}r}$ and $P$ (Equations~(\ref{eq:jumpxi}) and (\ref{eq:jumpp})) become
\begin{equation}
 [\xi_{{\rm i}r}] = - i \pi \frac{m^2/\ra^2}{\rhoi \left| \da \right|} P, \qquad \left[ P \right] = 0. \label{eq:jump0}
\end{equation}

Finally, the asymptotic expansion of function $\mathcal{F}$ for $\tau \to \pm \infty$ gives as the asymptotic behavior of $\xi_{{\rm i}\perp}$ when we move away from the resonance position as
\begin{equation}
 \xi_{{\rm i}\perp} \sim \frac{\mathcal{C}_{\rm A}}{\rhoi B} \frac{1}{\da \dima} \frac{i}{ \tau}, \label{eq:jumpxip}
\end{equation}
which again corresponds with the asymptotic dependence $\tau^{-1}$ found in both ideal and dissipative MHD.

In this Section we have studied the case of propagating waves, i.e., real $\omega$ and complex $k_z$. However, most of the analysis is equivalent for the case of standing quasi-modes, i.e., complex $\omega$ and real $k_z$. In particular, the jump conditions (Equations~(\ref{eq:jumpxi}) and (\ref{eq:jumpp})) are exactly the same for both propagating and standing modes. We refer the reader to, e.g., \citet{tirry} and \citet{vanlommel} for details about the derivation of the jump conditions for standing quasi-modes.

\section{Kink waves in straight tubes}
\label{sec:tube}

Here we perform an application of the theory of Section~\ref{sec:resonant} to the case of propagating resonant kink waves in straight flux tubes. 

We consider a straight magnetic cylinder of radius $R$ with a constant vertical magnetic field, i.e., $B_\varphi = 0$ and $B_z =$~constant. The tube is homogeneous in the longitudinal direction. In the radial direction, the tube is composed of an internal region with constant ion density $\rhoi_{\rm 1}$, a nonuniform transitional layer of thickness $l$, and an external region with constant ion density $\rhoi_{\rm 2}$. In the nonuniform layer the density changes continuously from the internal to the external values in the interval $R-l/2 < r < R+l/2$. Hereafter, indices `1' and `2' refer to the internal and external regions, respectively. For the sake of simplicity, the parameter $\alpha$ is taken constant everywhere so that the neutral density, $\rhon$, follows the radial dependence of the ion density. The ion-neutral collision frequency, $\nu_{\rm in}$, is also a constant to simplify matters.

We restrict ourselves to the case of thin tubes, i.e., the wavelength is much longer than the radius of the tube. For propagating waves the thin tube (TT) approximation is equivalent to the low-frequency approximation, i.e., $\omega R / \vk \ll 1$, with $\vk$ the kink velocity of ions defined as
\begin{equation}
\vk = \sqrt{\frac{\rhoi_{\rm 1} {v^2_{\rm Ai}}_{\rm 1} + \rhoi_{\rm 2} {v^2_{\rm Ai}}_{\rm 2}}{\rhoi_{\rm 1} + \rhoi_{\rm 2}}}.
\end{equation}
To check whether this approximation is realistic, let us consider the properties of the observed waves propagating in chromospheric spicules \citep[e.g.,][]{depontieu07, okamotodepontieu}. Taking $R=200$~km for the spicule radius and $\vk = 270$~km~s$^{-1}$ for the averaged phase velocity as estimated by \citet{okamotodepontieu}, we obtain  $\omega R / \vk \approx$~0.1 for a period of 45~s \citep{okamotodepontieu} and $\omega R / \vk \approx$~0.02 for a period of 3~min \citep{depontieu07}. Therefore, the use of the TT approximation is justified.  

 Additionally, we use the Thin Boundary (TB) approximation \citep[see, e.g.,][]{hollwegyang} and assume $l/R \ll 1$. The TB approximation enables us to use the jump conditions given in Equation~(\ref{eq:jump0})  as boundary conditions for the wave perturbations at the tube boundary. Detailed explanations of this method can be found in, e.g., \citet{goossens06} and \citet{goossens08}. Doing so, we arrive at the dispersion relation for resonant kink ($m=1$) and fluting ($m \geq 2$) waves in the TT and TB approximations. For simplicity we omit the intermediate steps which can be found in, e.g., \citet{goossens92}. The dispersion relation is
\begin{equation}
 \frac{1}{\rhoi_{\rm 1} \left( \omegat^2 - {\wa^2}_{\rm 1}  \right)} + \frac{1}{\rhoi_{\rm 2} \left( \omegat^2 - {\wa^2}_{\rm 2}  \right)} = i \pi   \frac{m/R}{\rhoi(R) \left| \da \right|_R}. \label{eq:disperthin}
\end{equation}
In Equation~(\ref{eq:disperthin})  we approximated $\ra \approx R$. The effect of ion-neutral collisions is enclosed in the definition of $\omegat$ (Equation~(\ref{eq:omegat})). For fixed and real $\omega$ the solution of Equation~(\ref{eq:disperthin}) is a complex longitudinal wavenumber, $k_z = k_{z \rm R} + i k_{z \rm I}$, where $k_{z \rm R}$ and $k_{z \rm I}$ are the real and imaginary parts of $k_z$, respectively. In the present normal mode analysis, which represents the stationary state of wave propagation, the amplitude of the propagating wave is proportional to $\exp \left( - z/\ld \right)$, with $\ld$ the damping length defined as
\begin{equation}
\ld = \frac{1}{k_{z \rm I}}.
\end{equation}

We investigate kink waves and set $m=1$.  In the absence of resonant absorption the right-hand side of Equation~(\ref{eq:disperthin}) is zero. In such a case, we can obtain an approximate expression for $k_{z \rm R}^2$ by assuming $k_{z \rm I}^2 \ll k_{z \rm R}^2$, namely
\begin{equation}
k_{z \rm R}^2 \approx \frac{\omega^2}{\vk^2} \frac{\omega^2+ \left( 1+\alpha \right) \nu_{\rm in}^2}{\omega^2+ \nu_{\rm in}^2}. \label{eq:kzr}
\end{equation}
In the single-fluid limit, $\nu_{\rm in} \gg \omega$ and $k_{z \rm R}^2 \approx \omega^2 \left( 1+\alpha \right) / \vk^2$, while in the fully ionized case $\alpha=0$ and $k_{z \rm R}^2 \approx \omega^2 / \vk^2$. Note that the expressions for $k_{z \rm R}^2$ in the single-fluid and fully ionized cases differ by a factor $\left( 1+\alpha \right)$. SOB studied propagating kink waves in the single-fluid approximation and found the same expression for $k_{z \rm R}^2$ for both partially ionized and fully ionized plasmas (see their Equations~(11) and (15)). The reason is that  SOB used a slightly different definition of the kink velocity, $\vk$. Here, the kink velocity is defined using the ion density only, while SOB defined $\vk$ using the total (ion + neutral) density. In SOB the total density is fixed and the effect of the ionization degree is to change the relative ion and neutral densities. On the contrary, here the ion density is a fixed parameter while the amount of neutrals depends on the ionization degree.

\subsection{Approximate expression for the damping length}

Here we seek an approximate expression for $\ld$. First we evaluate the factor $\rhoi(R) \left| \da \right|_R$ in Equation~(\ref{eq:disperthin}) by using the resonant condition $\omega^2 = \omega_{\rm A}^2 \left(  R \right)$, namely
\begin{equation}
\rhoi(R) \left| \da \right|_R = \omega^2 \left| \frac{{\rm d}\rhoi}{{\rm d}r} \right|_R,
\end{equation} 
where $\left| {\rm d}\rhoi / {\rm d}r \right|_R$ is the radial derivative of the density profile at $r = R$. Next we write $k_z = k_{z \rm R} + i k_{z \rm I}$ and assume weak damping, so we neglect terms with $\mathcal{O} \left( k_{z \rm I}^2 \right)$. An expression for the ratio $k_{z \rm I} / k_{z \rm R}$ is obtained from Equation~(\ref{eq:disperthin}), namely
\begin{equation}
\frac{k_{z \rm I}}{k_{z \rm R}} \approx \frac{1}{2} \frac{\alpha \omega \nu_{\rm in}}{\omega^2 + \left( 1 + \alpha \right) \nu_{\rm in}^2} + \frac{\pi}{8} \frac{1}{R} \frac{\left( \rhoi_{\rm 1} - \rhoi_{\rm 2} \right)^2}{\rhoi_{\rm 1} + \rhoi_{\rm 2}} \frac{1}{\left| {\rm d}\rhoi / {\rm d}r \right|_R}. \label{eq:ratiokz}
\end{equation}
Now we express $\left| {\rm d}\rhoi / {\rm d}r \right|_R$ as
\begin{equation}
\left| \frac{{\rm d}\rhoi}{{\rm d}r} \right|_R = F \frac{\pi^2}{4} \frac{\rhoi_{\rm 1} - \rhoi_{\rm 2}}{l},
\end{equation}
with $F$ a numerical factor that depends on the form of the density profile. For example, $F = 4/\pi^2$ for a linear profile \citep{goossens2002} and $F = 2/\pi$ for a sinusoidal profile \citep{rudermanroberts}. We assume that $k_{z \rm R}$ is approximately the same as in case without resonant damping (Equation~(\ref{eq:kzr})) and work on Equation~(\ref{eq:ratiokz}) to find  the expression for $\ld$ as, 
\begin{equation}
\frac{1}{\ld} \approx  \frac{1}{L_{\rm D, RA}} + \frac{1}{L_{\rm D, IN}}, \label{eq:ldgen}
\end{equation}
with  $L_{\rm D, RA}$ and $L_{\rm D, IN}$ the damping lengths due to resonant absorption and ion-neutral collisions, respectively, given by
\begin{eqnarray}
L_{\rm D, RA} &=& 2 \pi \mathcal{F} \vk \frac{R}{l} \frac{\zeta + 1}{\zeta - 1} \frac{1}{\omega} \left( \frac{\omega^2+\nu_{\rm in}^2}{\omega^2+ \left( 1+\alpha \right) \nu_{\rm in}^2} \right)^{1/2}, \label{eq:ldra}\\
L_{\rm D, IN} &=& 2 \vk \frac{\left( \omega^2 + \left( 1 + \alpha \right) \nu_{\rm in}^2 \right)^{1/2} \left( \omega^2+\nu_{\rm in}^2 \right)^{1/2}}{\alpha \omega^2\nu_{\rm in}} ,  \label{eq:ldin}
\end{eqnarray}
with $\zeta = \rhoi_1 / \rhoi_2$ the ion density contrast. Importantly, we find that both the collision frequency, $\nu_{\rm in}$, and the ionization degree, $\alpha$, are present in the expression of the damping length by resonant absorption (Equation~(\ref{eq:ldra})). In the single-fluid limit ($\nu_{\rm in} \gg \omega$) Equations~(\ref{eq:ldra}) and (\ref{eq:ldin}) become
\begin{eqnarray}
L_{\rm D, RA} &\approx & 2 \pi \mathcal{F} \vk \frac{R}{l} \frac{\zeta + 1}{\zeta - 1} \frac{1}{\omega} \left( 1+\alpha \right)^{-1/2}, \label{eq:ldrasf}\\
L_{\rm D, IN} &\approx & 2 \vk \frac{\left( 1+\alpha \right)^{1/2}  \nu_{\rm in}}{\alpha} \frac{1}{\omega^2}. \label{eq:ldinsf}\
\end{eqnarray}
The damping length due to resonant absorption in the single-fluid limit is inversely proportional to the frequency as in TGV. Indeed, in the fully ionized case ($\alpha=0$) Equations~(\ref{eq:ldra}) and (\ref{eq:ldrasf}) become Equation~(22) of TGV. For a partially ionized plasma $L_{\rm D, RA}$ also depends on the ionization degree, $\alpha$. The dependence on the ionization degree  was not discussed by SOB. SOB obtained an expression for the ratio of the damping length to the wavelength (see their Equation~(20)) and did not explicitly write the expression for the damping length. It turns out that the factor containing the ionization degree cancels out when the ratio of the damping length to the wavelength is computed, and so the dependence of $L_{\rm D, RA}$ on the ionization degree was not noticed by SOB. The damping length by ion-neutral collisions in the single-fluid limit (Equation~(\ref{eq:ldinsf})) is inversely proportional to $\omega^{2}$. This result is consistent with the expressions found by, e.g,  \citet{hearendel,pecseli,depontieu98}, and SOB among others.

\begin{figure}[!t]
\centering
 \includegraphics[width=0.95\columnwidth]{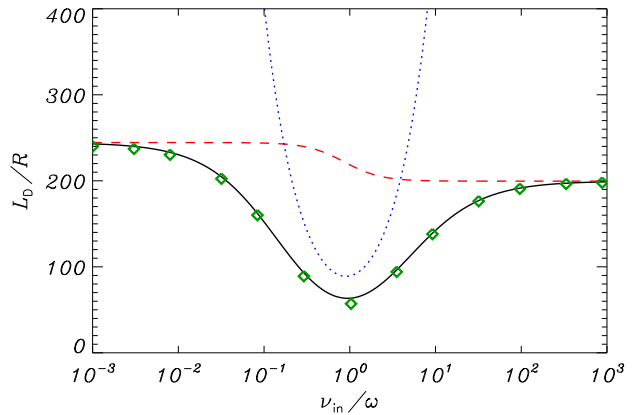}
\caption{$\ld/R$ vs. $\nu_{\rm in}/\omega$ for the kink wave in a magnetic flux tube with $l/R=0.2$, $\alpha=0.5$, $\zeta=10$, $F=2/\pi$, and $\omega R / \vk = 0.1$. The solid line is the total damping length in the TT and TB approximations computed from Equation~(\ref{eq:ldgen}). The dashed line is the damping length due to resonant absorption (Equation~(\ref{eq:ldra})) only, and the dotted line is the damping length due to ion-neutral collisions  (Equation~(\ref{eq:ldin})) only. Symbols $\Diamond$ correspond to the full numerical eigenvalue results.  \label{fig:ld}}
\end{figure}

Figure~\ref{fig:ld} shows $\ld/R$ versus $\nu_{\rm in}/\omega$ for a particular set of parameters given in the caption of the Figure. The total damping length (solid line) has a minimum for $\nu_{\rm in} \sim \omega$. The reason for this behavior is that damping by ion-neutral collisions (dotted line) becomes more relevant than resonant damping (dashed line) for $\nu_{\rm in} \sim \omega$.  In both the collisionless ($\nu_{\rm in} \ll \omega$) and single-fluid ($\nu_{\rm in} \gg \omega$) limits, the total damping length is well approximated by the damping length due to resonant absorption. In particular, for $\nu_{\rm in} \gg \omega$ the result tends to that predicted by Equation~(\ref{eq:ldrasf}).

\begin{figure}[!t]
\centering
 \includegraphics[width=0.95\columnwidth]{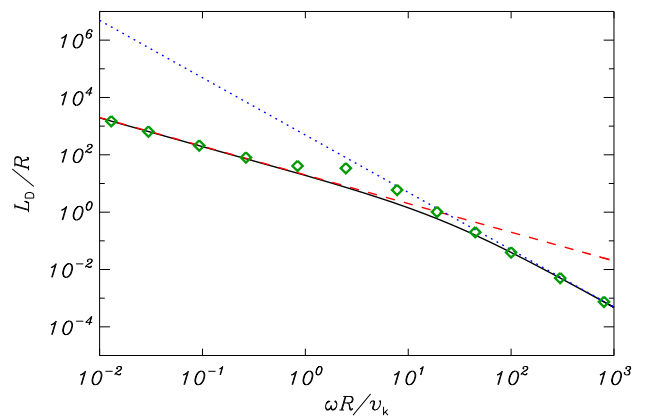}
\caption{$\ld/R$ vs. $\omega R / \vk $  with $\nu_{\rm in} R / \vk = 100$. The remaining parameters and the meaning of the line styles are the same as in Figure~\ref{fig:ld}.  \label{fig:ldw}}
\end{figure}

Now we fix the collision frequency to $\nu_{\rm in} R / \vk = 100$ and compute $\ld/R$ as a function of $\omega R / \vk $ (see Figure~\ref{fig:ldw}). The remaining parameters are the same as in Figure~\ref{fig:ld}. Although the observed wave frequencies in the solar atmosphere correspond to $\omega R / \vk \ll 1$, it is interesting to perform a general study for high values of the frequency. Our results point out that  the damping of the kink wave is governed by different mechanisms depending on the value of $\omega R / \vk $. We find that the damping length for $\omega R / \vk \ll 1$ is dominated by resonant absorption. Ion-neutral collisions start to become important when $\omega R / \vk \sim 10$. Finally, collisions are the dominant damping mechanism for large $\omega R / \vk$. We must note that the analytical solution in the TT approximation may not provide accurate results for values of $\omega R / \vk$ departing from the limit  $\omega R / \vk \ll 1$. For this reason, numerical eigenvalue computations which overcome the limitations of the analytical approximations are performed in the next Subsection.

\subsection{Numerical results}

Here we compare the approximate analytical result obtained by solving the dispersion relation in the TT and TB approximations (Equation~(\ref{eq:ldgen})) with the damping length obtained by solving the full eigenvalue problem numerically. The numerical solution is not limited by the TT and TB approximations. The numerical code is similar to that used in TGV and \citet{resonantflow}. The reader is refereed to these papers for details of the numerical scheme. In short, Equations~(\ref{eq:momion})--(\ref{eq:induct}) are integrated in the radial direction assuming a time dependence of the form $\exp \left( - i \omega t \right)$ and a spatial dependence in the $\varphi$ and $z$ directions as $\exp \left( i \varphi + i k_z z \right)$. The code solves the eigenvalue problem for the temporal damping of standing waves, i.e., complex $\omega$ provided a fixed and real $k_z$. To study spatial damping we need to convert the results from complex $\omega$ and real $k_z$ to real $\omega$ and complex $k_z$. The conversion is done following the method explained in TGV (see their Equation~(40)).

First, we consider the same parameters as in Figure~\ref{fig:ld} and compute the numerically determined $\ld/R$ versus $\nu_{\rm in}/\omega$. To compare with the analytical approximation, the numerical result is plotted using symbols $\Diamond$ in Figure~\ref{fig:ld}. A very good agreement between approximate and numerical results is found. This means that for $\omega R / \vk \ll 1$ the approximate analytical theory provides accurate results.

Next we numerically explore the effect of increasing the wave frequency. We take the same parameters as in Figure~\ref{fig:ldw}. Again, we use symbols $\Diamond$ to represent the eigenvalue result in Figure~\ref{fig:ldw}. We find that in the eigenvalue computations the transition between the regime dominated by resonant damping and that dominated by collisional damping occurs around $\omega R / \vk \sim 1$, while in  the analytical approximation collisions start to become important for $\omega R / \vk \sim 10$. This discrepancy is an effect of the TT approximation. For realistic values of the wave frequency ($\omega R / \vk \sim 10^{-2} - 10^{-1}$) both numerical and analytic results are in excellent agreement.

\section{Chromospheric kink waves}
\label{sec:chromos}

The results of Section~\ref{sec:tube} have direct implications for MHD waves propagating in partially ionized plasmas of the solar atmosphere. For kink waves studied in this paper, both resonant absorption and ion-neutral collisions decrease the amplitude of the waves. However, the two processes represent very different physical mechanisms.

On the one hand, resonant absorption is an ideal process that transfers wave energy from  global kink motions to localized azimuthal motions within the transversely inhomogeneous part of the flux tube. These azimuthal motions keep propagating along magnetic field lines \citep[see the numerical simulations by, e.g.,][]{pascoe1,pascoe2}. A detailed investigation of the energy transfer in the case of standing waves was done in \citet{arregui2d} by analyzing the Poynting flux in a two-dimensional configuration. However, resonant absorption itself does not dissipate wave energy in the plasma. The energy fed into the inhomogeneous layer will be dissipated by another mechanism later \citep[see, e.g., the results of][ in resistive MHD]{poedtskerner, poedtskerner2,poedts,poedts2,poedts3}. Hence, the damping length due to resonant absorption, $L_{\rm D, RA}$, represents the length scale for the kink motions to be converted into azimuthal motions.

On the other hand, ion-neutral collisions is a true dissipative process which deposits wave energy {\em in situ} and so it contributes to plasma heating. Hence, the damping length due to   ion-neutral collisions, $L_{\rm D, IN}$, represents the length scale for the kink wave energy to be dissipated by ion-neutral collisions. By comparing the values of $L_{\rm D, RA}$ and $L_{\rm D, IN}$ we can estimate the fraction of energy converted to Alfv\'enic, azimuthal motions and the fraction of energy dissipated by collisions.  

\begin{figure}[!t]
\centering 
\includegraphics[width=0.95\columnwidth]{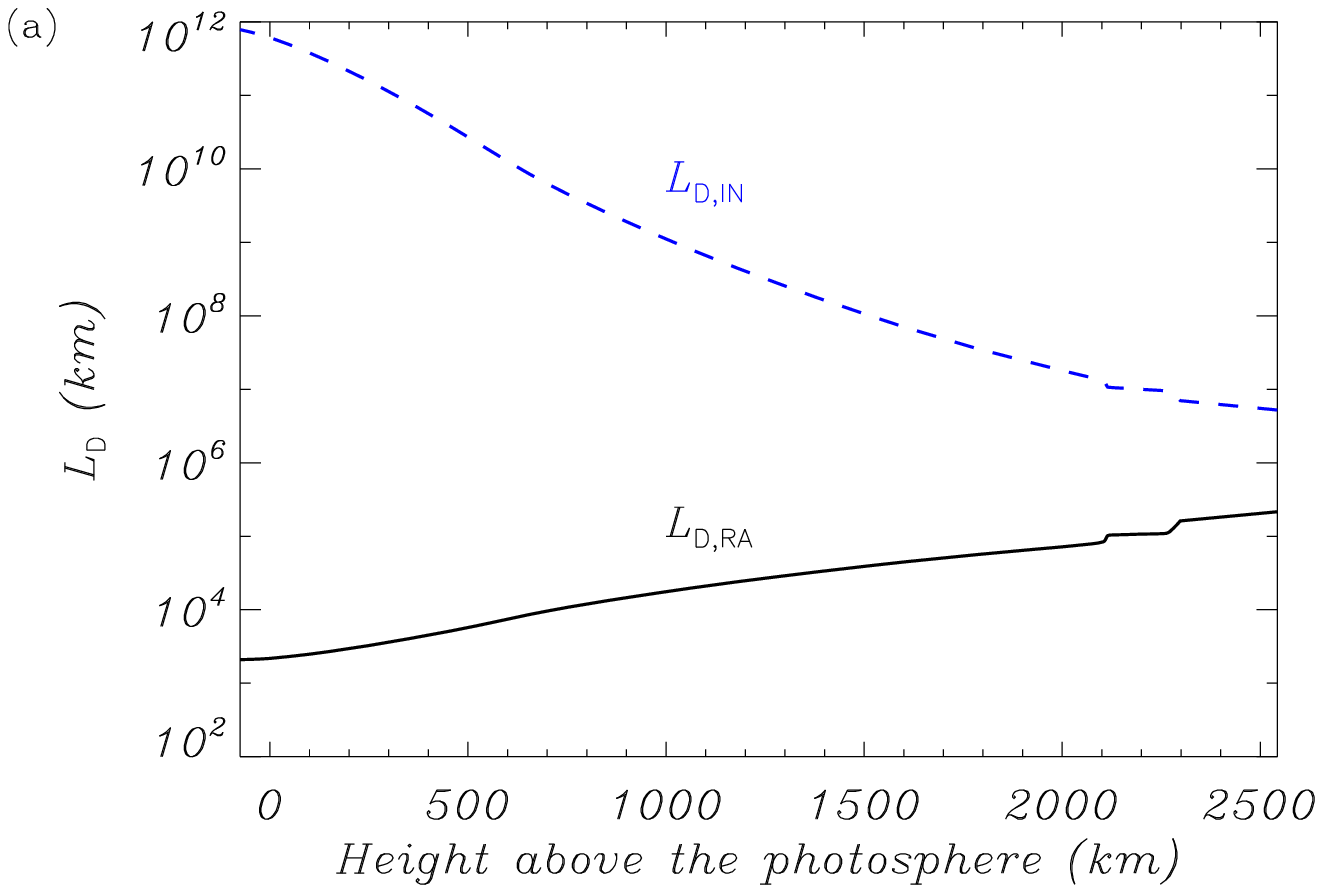}
 \includegraphics[width=0.95\columnwidth]{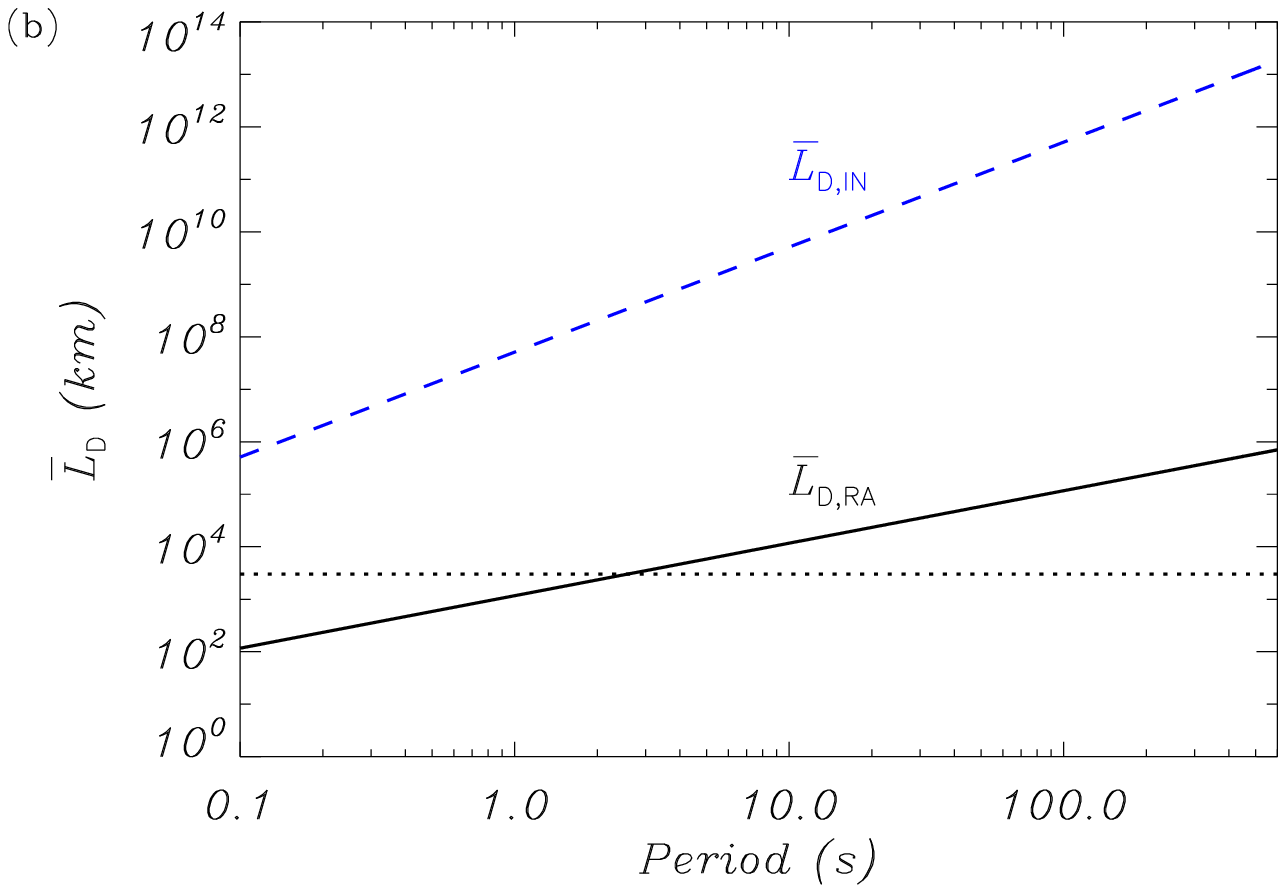}
\caption{(a) $L_{\rm D, RA}$ (solid line) and $L_{\rm D, IN}$ (dashed line)  vs. height above the photosphere for a wave period of 45~s. (b) Averaged $L_{\rm D, RA}$ (solid line) and $L_{\rm D, IN}$ (dashed line) in the chromosphere  vs. wave period. The horizontal dotted line represents the height of the chromosphere above the photosphere.  \label{fig:averagedld}}
\end{figure}

Let us apply this theory to kink waves propagating along chromospheric waveguides (spicules). We assume that the driver of the waves is located at the photosperic level and the waves propagate through the chromosphere to the corona. We take the variation of physical parameters with height (e.g., density, temperature, ionization degree, etc.) from the VALC model \citep{valc}. For the chromospheric magnetic field we consider the model used by \citet{leakearber}. Then we use Equations~(\ref{eq:ldrasf}) and (\ref{eq:ldinsf}) to compute the values of $L_{\rm D, RA}$ and $L_{\rm D, IN}$. As the physical parameters change along the spicule, both $L_{\rm D, RA}$ and $L_{\rm D, IN}$ are functions of height in the chromosphere.  We plot in Figure~\ref{fig:averagedld}(a) the values of the damping lengths as functions of height for a wave period of 45~s \citep{okamotodepontieu}. The damping length due to resonant absorption increases with height. This is an effect of the increase of the kink velocity, $\vk$, with height \citep{stratified}. At low heights, $L_{\rm D, RA}$ is comparable to the thickness of the whole chromosphere, meaning that a large fraction of wave energy is in the form of azimuthal motions when the wave reaches the coronal level. On the contrary, $L_{\rm D, IN}$ is several orders of magnitude longer. As $L_{\rm D, IN}$ decreases with height, the effect of collisions is more important in the upper chromosphere.

Next, we calculate the damping length averaged along the spicule, $\bar{L}_{\rm D}$, as
\begin{equation}
 \bar{L}_{\rm D} = \frac{1}{H} \int_0^H L_{\rm D} (s) {\rm d} s, \label{eq:averagedld}
\end{equation}
 where $s$ represents the direction along the spicule and $H$ is the height of the chromosphere above the photosphere. We take $H=$~3,000~km. Equation~(\ref{eq:averagedld}) is used to calculate the averaged values of both $L_{\rm D, RA}$ and $L_{\rm D, IN}$. We plot in Figure~\ref{fig:averagedld}(b) the averaged values of the damping lengths as functions of the wave period. First, we obtain that the averaged $L_{\rm D, IN}$ is several orders of magnitude longer than the averaged $L_{\rm D, RA}$ in the range of periods taken into account in Figure~\ref{fig:averagedld}(b). This means that ion-neutral collisions have little impact on wave propagation. On the contrary, the averaged $L_{\rm D, RA}$ for periods less than 10~s is smaller than or of the same order as the height of the chromosphere. This result points out that only waves with periods longer than 10~s are able to reach the coronal level in the form of kink motions. Waves with shorter periods reach the corona as small-scale azimuthal motions, which are unobservable with present day instruments. This effectively imposes a lower limit for the period of kink waves observable in the corona. This is consistent with the observed periods of coronal waves \citep[e.g.,][]{tomczyk07,tomczyk09,mcintosh2011}.

\section{Conclusion}

\label{sec:con}

In this paper we have investigated resonant Alfv\'en waves in partially ionized plasmas. We find that the conserved quantity at the resonance and the jump of the perturbations across the resonant layer are the same as in fully ionized, ideal plasmas. We have derived expressions for the damping lengths due to resonant absorption and due to ion-neutral collisions for the case of propagating kink waves in straight magnetic tubes. In the limit of large collision frequencies, the damping length due to resonant absorption is inverselly proportional to the frequency as in the fully ionized case \citep[see][]{TGV}, whereas the damping length due to ion-neutral collisions is inverselly proportional to the square of the frequency. We have applied the theory to the case of chromospheric kink waves propagating from the photosphere to the corona. We conclude that the solar chromosphere acts as a filter for kink waves. Waves with peridos shorter than  10~s, approximetely, reach the corona in the form of small-scale azimuthal waves. Only waves with periods longer than 10~s can be  observed in the corona as kink waves.

\acknowledgements{
  We thank I. Arregui and A. J. D\'iaz  for useful comments. R.S. acknowledges support from a Marie Curie Intra-European Fellowship within the European Commission 7th Framework Program  (PIEF-GA-2010-274716).  M.G. acknowledges support from K.U. Leuven via GOA/2009-009}

\end{document}